# R-process enrichment from a single event in an ancient dwarf galaxy

Alexander P. Ji[1], Anna Frebel[1,2], Anirudh Chiti[1], Joshua D. Simon[3]

**The elements heavier than zinc are synthesized through the (r)apid and (s)low neutron-capture processes[1,2]. The primary astrophysical production site of the r-process elements (such as europium) has been debated for nearly 60 years[2]. Chemical abundance trends of old Galactic halo stars initially suggested continual r-process production from sources like core-collapse supernovae[3,4], but evidence in the local universe favored r-process production primarily from rare events like neutron star mergers[5,6]. The appearance of a europium abundance plateau in some dwarf spheroidal galaxies was suggested as evidence for rare r-process enrichment in the early universe[7], but only under the assumption of no gas accretion into the dwarf galaxies. Invoking cosmologically motivated gas accretion[8] actually favors continual r-process enrichment in those systems. Furthermore, the universal r-process pattern[1,9] has not been cleanly identified in dwarf spheroidals. The smaller, chemically simpler, and more ancient ultra-faint dwarf galaxies assembled shortly after the formation of the first stars and are ideal systems to study nucleosynthesis processes such as the r-process[10,11]. Reticulum II is a recently discovered ultra-faint dwarf galaxy[12-14]. Like other such galaxies, the abundances of non-neutron-capture elements are similar to those of other old stars[15]. Here we report that seven of nine stars in Reticulum II observed with high-resolution spectroscopy show strong enhancements in heavy neutron-capture elements with abundances that exactly follow the universal r-process pattern above barium. The enhancement in this "r-process galaxy" is 2-3 orders of magnitude higher than what is seen in any other ultra-faint dwarf galaxy[11,16,17]. This implies that a single rare event produced the r-process material in Reticulum II, whether or not gas accretion was significant in ultra-faint dwarf galaxies. The r-process yield and event rate is incompatible with ordinary core-collapse supernova[18] but consistent with other possible sites, such as neutron star mergers[19].**

Ultra-faint dwarfs (UFDs) are small galaxies that orbit the Milky Way and have been discovered by deep wide-area sky surveys[12,13]. Although physically close to us, they are also relics from the era of the first stars and galaxies and thus an ideal place to investigate the first metal enrichment events in the universe[10]. Observations of UFDs provide evidence that they form all their stars within 1-3 Gyr of the Big Bang[20], their stars contain very small amounts of elements heavier than helium ("metals")[21], and they are enriched by the metal output of only a few generations of stars[11,20]. The chemical abundances of light elements (less heavy than iron) suggested that core-collapse supernovae were the primary metal sources in these systems[11,16,17]. This conclusion was supported by unusually low neutron-capture element abundances that are consistent with small amounts of neutron-capture element production associated with massive star evolution[11]. Neutron star mergers may be the dominant source of r-process elements in the present day[5,6], but they were thought to be irrelevant in the low metallicity regime, including UFDs. The r-process yield was thought to be too high to be consistent with metal-poor stars[3], the occurrence rate of these binaries was too low to be found in typical UFDs[4,7], and the long merging delay time precluded significant contributions from neutron star mergers at early times[4,22,23].

---

[1] Dept. of Physics & Kavli Institute for Astrophysics and Space Research, Massachusetts Institute of Technology, Cambridge, MA 02139, USA
[2] Joint Institute for Nuclear Astrophysics, Center for the Evolution of the Elements, East Lansing, MI 48824, USA
[3] Observatories of the Carnegie Institution of Washington, Pasadena, CA 91101, USA

The UFD Reticulum II (Ret II) was recently discovered with Dark Energy Survey data[12,13] and confirmed to be one of the most metal-poor galaxies known[14]. On 1-4 Oct 2015, we obtained high-resolution spectra of the nine brightest member stars in Ret II (see Table 1, Extended Data Figure 1). The abundances of non-neutron-capture elements in all nine stars are consistent with abundances in other UFD stars and Galactic halo stars[11,15,24]. Surprisingly, only the two most metal-poor stars show the deficiency of neutron-capture elements typically found in UFDs[11,16], whereas the remaining seven stars display extremely strong spectral lines of europium and other neutron-capture elements (Figure 1). Our abundance analysis (see Methods) finds that these seven stars span a factor of 10 in metallicity centered at [Fe/H] = –2.5, and all seven stars are significantly enhanced in neutron-capture elements. Their [Eu/Fe] abundances are the highest found in any dwarf galaxy so far[7,16] and are comparable to the most Eu-enhanced halo stars currently known[24] (Figure 2a,b). Surface accretion of neutron-capture elements from the interstellar medium is at least 1,000x too small to account for this level of enhancement[25], and binary mass transfer is unlikely to enrich multiple stars in this galaxy. Thus, these stars formed from gas that was heavily pre-enriched with neutron-capture elements.

From our spectra of the brightest four neutron-capture-rich stars, we measure 3-8 additional Rare Earth element abundances. The relative abundances of elements with atomic number >55 unambiguously match the scaled solar r-process pattern[1,9] (Figure 2c). The [Eu/Ba] ratios of the three fainter neutron-capture-rich stars also point to an r-process origin. Ret II thus appears to be an "r-process galaxy", with 78% of observed stars being highly enriched in r-process elements. In comparison, the frequency of metal-poor stars in the Milky Way halo with similar r-process enhancement is <5%[24]. Furthermore, all stars in the nine other UFDs with high-resolution neutron-capture abundances have [Ba/Fe] and [Eu/Fe] values at least 100 times lower than the stars in Ret II, although some of those UFDs currently have few stars with such measurements[11,16,17]. It is thus extremely likely that the neutron-capture material in Ret II was produced by just one event. If each UFD was equally likely to host an r-process event, then the probability that $N$ r-process events occurred in Ret II but zero r-process events occurred in the other nine UFDs is $(1/10)^N$. There is then only a ~1% chance that 2 or more events contributed to the r-process material in Ret II. While gas accretion could potentially hide a prolific r-process event in one of the other UFDs, accreting enough gas to decrease the neutron-capture abundance by over two orders of magnitude while leaving no stars with intermediate r-process enhancements is implausible.

The r-process yield of a typical core-collapse supernova cannot explain the high r-process abundances found in this galaxy. Using Eu as the representative r-process element, five of the r-process stars in Ret II have [Eu/H] of –1 to –1.3, implying these stars formed in an environment where the Eu mass ratio $M_{Eu}/M_H$ was $10^{-10.3}$ to $10^{-10.6}$. The two faintest r-process stars have higher [Eu/H] values, but their larger abundance uncertainties place them within 1-2σ of [Eu/H]= –1. In a UFD, metals are typically diluted into ~$10^6$ solar masses of hydrogen by turbulent mixing during galaxy assembly[10,26,27], with low and high limits of $10^5$-$10^7$ solar masses (see Methods). The Eu yield of this r-process event would then be $10^{-4.3}$ to $10^{-4.6}$ solar masses, 1,000x higher than typical core-collapse supernova yields ($M_{Eu}$~$10^{-7.5}$ solar masses[17], brown vertical bar in Figure 2a,b). Extreme supernova Eu yields of ~$10^{-6}$ solar masses have been previously invoked to aid in chemical evolution models[4], but even combining these with the minimum possible dilution mass results in [Eu/H] values too low to match the stellar abundances in Ret II.

Though several candidates exist for rare and prolific r-process sites, neutron star mergers are considered one of the most likely alternatives to supernovae[3-6,23,25-28]. Typical neutron star merger Eu yields are $M_{Eu}$ ~$10^{-4.5}$ solar masses[19], resulting in [Eu/H] values compatible with the abundances observed in Ret II (orange vertical bar in Figure 2a,b). The rate of this event also appears consistent with a neutron star merger, although both the observed and expected rates are uncertain (see Methods). Since only one prolific r-process event occurred in the ten UFDs observed so far, combining the present-day stellar mass of these ten UFDs allows an estimate of how many supernovae explode for one such event to occur. With a standard initial mass function, we find ~2,000 supernovae contributed to all ten of these UFDs, comparable to the expected average rate of one neutron star merger every 1,000-2,000 supernovae[7].

Chemical evolution models that incorporate yields of neutron star mergers typically need to invoke unusually short merging times of ~1 Myr to explain the observed halo star [Eu/Fe] and [Fe/H] distributions[4,23,28]. However, recent studies have found that a combination of inhomogeneous mixing, hierarchical galaxy formation, and inefficient star formation can alleviate this issue[25-28]. A neutron star merger in a UFD naturally produces such conditions. In particular, supernova feedback was especially effective at disrupting the small minihalo progenitors of UFDs. The resulting >10-100Myr delays between star formation episodes[10,29] are consistent with the shortest delay times predicted for neutron star mergers[22].

Our observations are also consistent with other rare and prolific r-process events. In particular, magnetorotationally driven supernovae have many desirable properties, synthesizing as much as ~$10^{-5}$ solar masses of Eu on a supernova timescale and at a rate more frequent than that of neutron star mergers[23]. These supernovae and neutron star mergers have sufficiently similar yields and rates that Ret II cannot yet be used to firmly distinguish between the two cases. If future theoretical work finds these two sites differ in other ways, such as the abundance of neutron-capture elements around the first r-process peak[15], the stellar abundances in Ret II may be used to eventually differentiate between them.

Previous evidence for rare and prolific r-process events in more luminous dwarf spheroidals relied on interpreting a flat [Eu/H] trend with respect to [Fe/H][7]. This plateau only favors a rare r-process event if gas accretion is insignificant in the galaxy. Invoking gas accretion actually lowers [Eu/H] with increasing metallicity, in which case the observed plateau requires continual Eu production from core-collapse supernovae rather than a single r-process event. We note that hierarchical structure formation predicts significant gas accretion into these larger dwarf galaxies, and extra gas is needed to reproduce their overall metallicity distribution functions[8].

In contrast, the evidence for a single event in Ret II is based on large [Eu/Fe] and [Ba/Fe] enhancements relative to those measured in stars in the other UFDs. The Eu and Ba trends within Ret II can then be used to understand the star formation, gas accretion, and metal mixing history of this galaxy. As an illustration, the highest [Fe/H] star may indicate the presence of inhomogeneous metal mixing as it has similar [Eu/Fe] compared with the lower [Fe/H] stars[30], though we caution that the current data give statistically insignificant abundance trends (see Methods). The stellar abundances in Ret II thus not only show that rare and prolific r-process events existed in the early universe, but they also hold the key to understanding the formation history of this relic from the era of first galaxies.

**Acknowledgments** Our data were gathered using the 6.5-m Magellan Clay telescope located at Las Campanas Observatory, Chile. A.P.J. thanks N. Weinberg and P. Schechter for discussions. A.P.J. and A.F. are supported by NSF-CAREER grant AST-1255160. A.F. acknowledges support from the Silverman (1968) Family Career Development Professorship. J.D.S. acknowledges support from NSF grant AST-1108811. This work made use of NASA's Astrophysics Data System Bibliographic Services and the open-source python libraries numpy, scipy, matplotlib, statsmodels, pandas, seaborn, and astropy.

This research made use of data products originally obtained with the Dark Energy Camera (DECam), which was constructed by the Dark Energy Survey (DES) collaboration. Funding for the DES Projects has been provided by the DOE and NSF (USA), MISE (Spain), STFC (UK), HEFCE (UK). NCSA (UIUC), KICP (U. Chicago), CCAPP (Ohio State), MIFPA (Texas A&M), CNPQ, FAPERJ, FINEP (Brazil), MINECO (Spain), DFG (Germany) and the collaborating institutions in the Dark Energy Survey, which are Argonne Lab, UC Santa Cruz, University of Cambridge, CIEMAT-Madrid, University of Chicago, University College London, DES-Brazil Consortium, University of Edinburgh, ETH Zürich, Fermilab, University of Illinois, ICE (IEEC-CSIC), IFAE Barcelona, Lawrence Berkeley Lab, LMU München and the associated Excellence Cluster Universe, University of Michigan, NOAO, University of Nottingham, Ohio State University, University of Pennsylvania, University of Portsmouth, SLAC National Lab, Stanford University, University of Sussex, and Texas A&M University.

**Author Contributions** A.P.J. took the observations and led the analysis and paper writing; A.F. and A.C. assisted with the observations; A.F. and J.D.S. contributed to the analysis; all authors contributed to writing the paper.

**Author Information** Reprints and permissions information is available at www.nature.com/reprints. The authors declare no competing financial interests. Correspondence and requests for materials should be addressed to A.P.J. (alexji@mit.edu).

**Table 1: Properties of the nine observed Reticulum II stars.**

| RA | DEC | $v_{hel}$ | $g$ | $T_{eff}$ | log g | $v_t$ | [Fe/H] | [Ba/Fe] | [Eu/Fe] |
|---|---|---|---|---|---|---|---|---|---|
| 3h35m23.85s | −54d04m07.50s | 66.8 | 16.45 | 4608 | 1.00 | 2.40 | −3.01 | 0.79 | 1.68 |
| 3h36m07.75s | −54d02m35.56s | 62.7 | 17.43 | 4833 | 1.55 | 2.15 | −2.97 | 0.91 | 1.74 |
| 3h34m47.94s | −54d05m25.01s | 62.0 | 17.52 | 4900 | 1.70 | 1.90 | −2.91 | 1.08 | 1.87 |
| 3h35m31.14s | −54d01m48.25s | 60.9 | 17.64 | 4925 | 1.90 | 1.80 | −3.34 | <−0.80 | <1.50 |
| 3h35m48.04s | −54d03m49.82s | 61.9 | 18.27 | 5125 | 2.35 | 1.75 | −2.19 | 0.36 | 0.95 |
| 3h35m37.06s | −54d04m01.24s | 63.5 | 18.57 | 5170 | 2.45 | 1.55 | −2.73 | 1.40 | 1.70 |
| 3h35m56.28s | −54d03m16.27s | 62.7 | 18.85 | 5305 | 2.95 | 1.65 | −3.54 | < 0.10 | <2.40 |
| 3h34m57.57s | −54d05m31.42s | 61.9 | 18.94 | 5328 | 2.85 | 1.50 | −2.08 | 1.36 | 1.77 |
| 3h34m54.24s | −54d05m58.02s | 71.6 | 18.95 | 5395 | 3.10 | 1.35 | −2.77 | 1.40 | 2.11 |

Right Ascension (RA) and Declination (DEC) indicate star coordinates. $v_{hel}$ is the heliocentric radial velocity in km s$^{-1}$. $g$ is the star's magnitude. Stellar parameters are effective temperature ($T_{eff}$ [K]), surface gravity (log g [dex]), and microturbulence ($v_t$ [km s$^{-1}$]). The notation [A/B] = $\log_{10}(N_A/N_B)_{star} - \log_{10}(N_A/N_B)_{sun}$ quantifies the logarithmic number ratio between two elements A and B relative to the solar ratio.

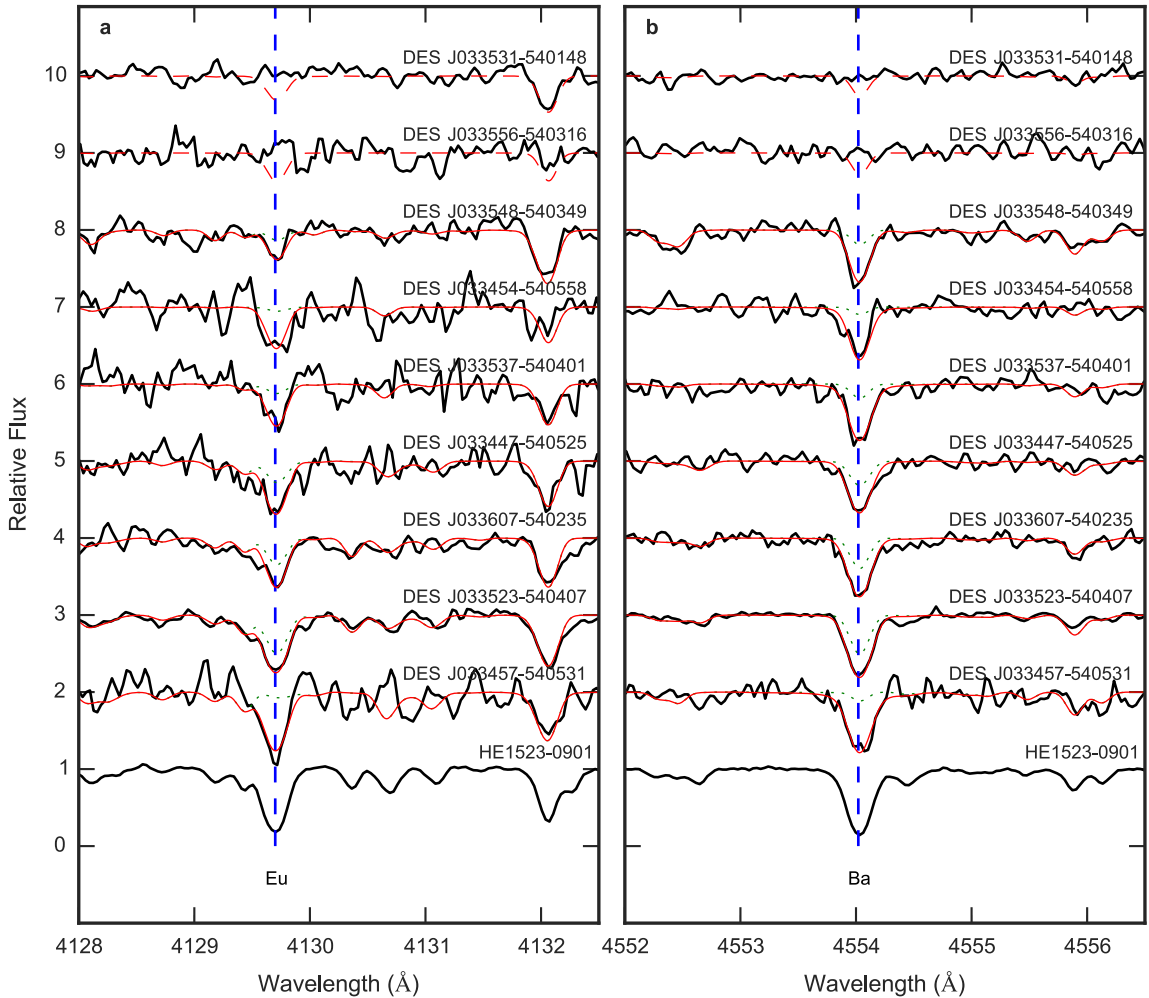

**Figure 1:** Spectra of stars in Reticulum II
**a**, Region around Eu absorption line (412.9 nm). Absorption is clearly present in seven of the nine Ret II spectra (black lines), including those with modest signal-to-noise. Thin red lines show synthesized fits to the absorption lines (dashed lines for upper limits). For comparison, dotted green lines are synthesized spectra for each individual star using typical limits found in other UFDs ([Eu/H]= –2.0). Also shown is HE1523–0901, one of the most r-process enhanced halo stars known[1]. **b,** Same as **a** but around Ba absorption line (455.4 nm) ([Ba/H]= –4.0 in dotted green).

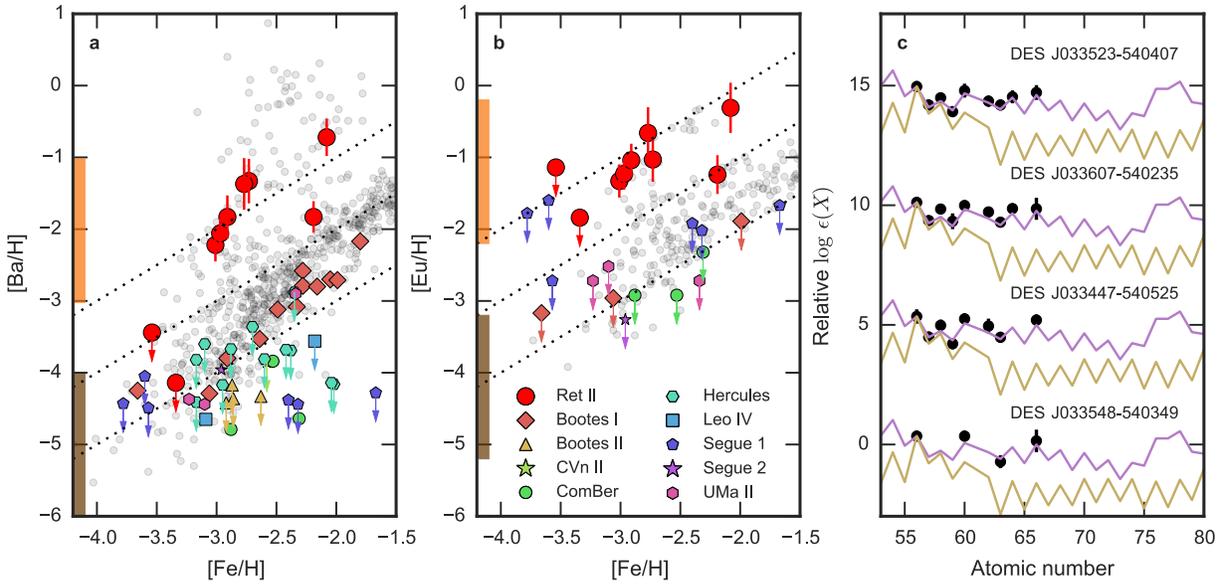

**Figure 2:** Chemical abundances of stars in Reticulum II
**a**, [Ba/H] and [Fe/H] of stars in Ret II (red points), in the halo[24] (gray points), and in UFDs (colored points, references within refs. 16, 17). Orange and brown vertical bars indicate expected abundance ranges following a neutron star merger and core-collapse supernova, respectively. Dotted black lines show constant [Ba/Fe]. Arrows denote upper limits. Error bars are 1σ (see Extended Data Table 1 and Methods). **b,** Same as **a** but for Eu.
**c**, Abundance patterns above Ba for the four brightest Eu-enhanced stars in Ret II (Extended Data Table 2), compared to solar r- and s-process patterns[9] (purple and yellow lines, respectively). Solar patterns are scaled to stellar Ba. Stars are offset by multiples of 5.

**Methods**
**Observations and Abundance Analysis**
The brightest known red giant members of Ret II were selected for observation from published medium-resolution spectroscopic surveys of Ret II[14,31,32] (Extended Data Figure 1). On 1-4 Oct 2015, we obtained high-resolution spectra with the MIKE spectrograph[33] on the Magellan-Clay telescope using a 1.0-arcsec slit and covering 3500Å to 9000Å. This provides a spectral resolution R of ~22,000 and ~28,000 at redder and bluer optical wavelengths, respectively. Stars were each observed for 1-4 hours, resulting in signal-to-noise per pixel of ~8-22 at 4250Å and 15-45 at 6000Å (Extended Data Table 1). We used the Carnegie-Python pipeline to reduce the spectra[34]. Separate echelle orders were normalized and summed with the semi-automated code SMH[35], which was also used for the abundance analysis. Radial velocities were determined by cross-correlation of the Mg triplet lines near 5150A against a high signal-to-noise spectrum of HD140283.

Stellar parameters and abundances were determined with standard spectroscopic methods[36], which we summarize briefly. Equivalent widths of iron lines were determined by fitting Gaussian profiles. We rejected iron lines whose reduced equivalent width is larger than –4.5 as these lines are likely past the linear regime of the curve of growth. We used the Castelli-Kurucz stellar atmospheres with enhanced alpha-abundances[37] and the 1D LTE code MOOG[38] to determine abundances of these lines. The effective temperature was found by requiring no iron abundance trend with respect to excitation potential. The surface gravity was found by requiring the Fe II lines to have the same abundance as the Fe I lines. For the 7$^{th}$ and 9$^{th}$ stars in Table 1, we used an isochrone[39] to determine the surface gravity as no Fe II lines were detectable. The microturbulence was found by requiring no trend between abundance and reduced equivalent width. After determining the stellar parameters spectroscopically, we applied an effective temperature correction[36] and redetermined the surface gravity and microturbulence. This correction increases the effective temperatures and results in increased metallicities.

Statistical uncertainties in the effective temperature and microturbulence were estimated by varying the parameters to match the standard deviation of the fitted slopes. Uncertainty in the surface gravity was estimated by varying the parameter to match the standard error of the Fe I and Fe II abundance. We adopt systematic stellar parameter uncertainties of 150K, 0.3 dex, and 0.2 km s$^{-1}$ for the effective temperature, surface gravity, and microturbulence respectively. These are added in quadrature to the 1σ statistical error. For surface gravity, we adopt 0.4 dex total uncertainty when no Fe II lines are available. The uncertainties are listed in Extended Data Table 1. The stellar parameter uncertainties typically correspond to a ~0.2-0.3 dex total uncertainty in iron abundance, which is dominated by the uncertainty in the effective temperature. The abundances of the brightest four stars have been confirmed by observations with higher signal-to-noise[15].

Abundances of neutron-capture elements were determined with a line list compiled from several sources (refs. 40, 41, and references in refs. 42, 43). We used spectrum synthesis to derive abundances of Ba, La, Pr, and Eu. Other neutron-capture element abundances were determined using equivalent widths of unblended lines. The abundances are tabulated in Extended Data Table 2. Abundance uncertainties indicate the larger of 1) the standard deviation of abundances derived from individual lines accounting for small-number statistics[44]; and 2) the total [Fe/H] uncertainty including stellar parameter uncertainties. The latter uncertainty typically dominates (see Extended Data Table 1). Abundances are quoted relative to solar abundances[45]. Conservative upper limits were determined by synthesizing a line with amplitude two times larger than the typical continuum uncertainty.

The strong 4554A and 4934A Ba lines are significantly affected by the isotope ratios of these lines. We have used the r-process only isotope ratios, which reduce the abundance derived from these lines by 0.1-0.3 dex compared to employing the solar Ba isotope ratios[1]. We also use the 5853A, 6141A, and 6496A lines to determine the Ba abundances. For Eu, we use the r-process only isotope fractions and measure the 4129A, 4205A, 4435A, 4522A, and 6645A lines. Full line lists can be obtained by contacting the authors.

**Dilution Mass for Ultra-faint Dwarf Galaxies**
To determine the yield of the r-process event, the observed ratio between Eu and H must be converted into a Eu mass. This requires finding the mass of hydrogen gas that the r-process material is diluted into. The dominant physical process affecting this dilution mass is turbulent mixing driven by the gravitational assembly of the UFD[10,46]. Cosmological simulations have not yet resolved this at the scale of individual metal enrichment events[27], but order-of-magnitude estimates of a turbulent diffusion coefficient and mixing time result in typical dilution masses of $\sim 10^6$ solar masses[10].

Stringent upper and lower bounds can be placed on the dilution mass. The total halo mass of an assembling UFD is $10^7$-$10^8$ solar masses, including dark matter[10]. With the cosmological baryon fraction, this places an upper bound of $\sim 10^7$ solar masses of hydrogen gas that is available to dilute into. A lower bound can be derived by assuming the limit of no turbulent mixing. For a $10^{51}$ erg supernova, this corresponds to $\sim 10^5$ solar masses[10]. For a $10^{50}$ erg neutron star merger, this limit is instead $10^{3.5-4}$ solar masses[27]. The estimate of $10^6$ solar masses of gas thus cannot be off by more than 1 order of magnitude in either direction.

Including a ±1 order of magnitude range on the mixing mass still rules out ordinary core-collapse supernovae as the source of the r-process material in Ret II. See Figure 2a,b, where the small shaded bars on the left show the [Ba/H] and [Eu/H] derived using a mixing mass of $10^5$ to $10^7$ solar masses of hydrogen and a fiducial yield. The adopted Eu yields are $10^{-7.5}$ solar masses for core-collapse supernovae[18] and $10^{-4.5}$ solar masses for neutron star mergers[19]. The Ba yield is calculated from the Eu yield assuming the r-process ratio[9] such that [Ba/Eu] = –0.82. Note that many dwarf galaxy stars in Figure 2b do not have abundances or upper limits for Eu. It is common to not report Eu upper limits when they are too large to be a relevant constraint.

**Expected Event Rate**
We derive an expected event rate by estimating the total number of supernovae that have exploded in all ten UFDs considered in this paper. The combined present day luminosity of these UFDs is $\sim 10^5$ $L_{sun}$[14,47]. Note that 80% of the stellar mass comes from just three galaxies (Hercules, Boo I, Leo IV).

Consider an initial mass function $\phi(m) \sim m^{-\alpha}$ with lower and upper mass limits $M_l$ and $M_u$. Let the minimum mass for a supernova be $M_{SN}$, and the maximum mass of a star that lives for the age of the universe be $M_{max}$. Then given a present day luminosity $L_0$ and mass-to-light ratio $\eta$, it is straightforward to show that the number of supernovae per surviving solar mass is:

$$\frac{\int_{M_{SN}}^{M_u} \phi(m)dm}{\int_{M_l}^{M_{max}} m\,\phi(m)dm}\eta L_0$$

Typical values are $M_l \sim 0.1$ $M_{sun}$, $M_u \sim 50$ $M_{sun}$, $M_{SN} \sim 10$ $M_{sun}$, $M_{max} \sim 0.8$ $M_{sun}$. We calculate $\eta=2.2$ with a Dartmouth isochrone (12Gyr, [Fe/H]=-2.5, [α/Fe]=0.4)[48]. The standard Salpeter initial mass function has $\alpha=2.35$, which results in the number of supernovae being $\sim 0.009 \eta L_0$, or $\sim 2,000$ supernovae from $\sim 10^5$ $L_{sun}$ of stars surviving until today. This rate is consistent with the average expected rate of neutron star mergers[7].

The agreement between the rates is promising but by no means conclusive, as several additional factors may affect the estimated rate. If these galaxies lost stellar mass due to tidal stripping, the number of supernovae would have been correspondingly larger. If the initial mass function in UFDs is not Salpeter, this will affect the number of supernovae as well. Indeed, the observed initial mass function for stars with M < 0.8 $M_{sun}$ in UFDs is bottom-light with a slope $\alpha \sim 1.3$ and mass-to-light ratio $\eta=0.92$. Though it is unknown whether the slope can be extrapolated to higher masses[11,49], such an extrapolation results in much larger numbers of supernovae. For the typical mass limits considered here, there are $0.683 \eta L_0$ supernovae, or 63,000 supernovae in all the UFDs combined. Additionally, the r-process in Ret II occurred at very low metallicity, perhaps implying a higher rate for the event than calculated here, because r-process elements synthesized by neutron star mergers occurring after star formation finished cannot be preserved in a galaxy's chemical abundances. Some UFDs may contain r-process enhanced stars not yet observed. The low-metallicity environment may also affect the binary fraction and merging delay time. We note that the expected rate of neutron star mergers is uncertain to 1-2 orders of magnitude even in the local universe[50]. A neutron star binary may experience large velocity kicks during its formation, which can eject the binary from its host galaxy and further reduce the expected rate of these mergers[51].

**Statistical Significance of Abundance Trends**
The [Ba/H] and [Eu/H] values visually appear to increase with [Fe/H]. However, we caution that these trends should not be over-interpreted due to the small number of stars and significant abundance uncertainties. As a simple illustration, we perform a weighted least squares regression (weighting by the inverse error) on the seven r-process stars and use a t-test to determine the significance of the slope. The [Ba/H] vs [Fe/H] abundance trend has a slope of 0.890, standard error of 0.446, and p-value of 0.10. The [Eu/H] vs [Fe/H] abundance trend has a slope of 0.486, standard error of 0.354, and a p-value of 0.23. Neither abundance correlation is statistically significant.

**Code Availability**
All codes used to reduce and analyze the data are publicly available. This includes the Carnegie-Python MIKE reduction pipeline[34] and the abundance analysis code MOOG[38].

| Star | S/N 4250A | N Fe I Lines | Statistical Uncertainty | | | Total Uncertainty | | | [Fe/H] total error | Ba error s.d. (N) | Eu error s.d. (N) |
|---|---|---|---|---|---|---|---|---|---|---|---|
| | | | $T_{eff}$ | log g | $v_t$ | $T_{eff}$ | log g | $v_t$ | | | |
| 1 | 22 | 128 | 46 | 0.02 | 0.21 | 157 | 0.30 | 0.29 | 0.23 | 0.06 (5) | 0.13 (5) |
| 2 | 12 | 103 | 71 | 0.16 | 0.19 | 166 | 0.34 | 0.28 | 0.21 | 0.17 (5) | 0.22 (4) |
| 3 | 11 | 104 | 81 | 0.08 | 0.20 | 170 | 0.31 | 0.28 | 0.23 | 0.30 (5) | 0.06 (3) |
| 4 | 16 | 80 | 63 | 0.19 | 0.20 | 163 | 0.36 | 0.28 | 0.23 | | |
| 5 | 12 | 124 | 62 | 0.12 | 0.20 | 162 | 0.32 | 0.28 | 0.22 | 0.11 (5) | 0.27 (2) |
| 6 | 10 | 51 | 134 | 0.21 | 0.30 | 201 | 0.37 | 0.36 | 0.31 | 0.30 (5) | 0.00 (2) |
| 7 | 11 | 33 | 210 | N/A | 0.35 | 258 | 0.40 | 0.40 | 0.37 | | |
| 8 | 8 | 67 | 104 | 0.11 | 0.23 | 183 | 0.32 | 0.30 | 0.26 | 0.04 (5) | 0.35 (3) |
| 9 | 7 | 31 | 199 | N/A | 0.37 | 249 | 0.40 | 0.42 | 0.36 | 0.24 (4) | 0.00 (2) |

**Extended Data Table 1:** Stellar parameter uncertainties
S/N is signal-to-noise ratio per pixel near 4250Å. $T_{eff}$ in K, log g in dex, $v_t$ in km s$^{-1}$. The number of lines used to determine the Ba and Eu abundances is given in parentheses. The adopted abundance error is the larger of the standard deviation between lines and the [Fe/H] error. Stars are numbered from brightest to faintest, from Star 1 to Star 9 in this order: DES J033523–540407, DES J033607–540235, DES J033447–540525, DES J033531–540148, DES J033548–540349, DES J033537–540401, DES J033556–540316, DES J033457–540531, DES J033454–540558

| Element | Star 1 | Star 2 | Star 3 | Star 4 | Star 5 | Star 6 | Star 7 | Star 8 | Star 9 |
|---|---|---|---|---|---|---|---|---|---|
| log ε(Ba) | −0.04 | 0.12 | 0.35 | <−1.96 | 0.35 | 0.85 | <−1.26 | 1.46 | 0.81 |
| [Ba/Fe] | 0.79 | 0.91 | 1.08 | <−0.80 | 0.36 | 1.40 | <0.10 | 1.36 | 1.40 |
| log ε(La) | −0.81 | −0.64 | −0.51 | | | | | | |
| [La/Fe] | 1.10 | 1.23 | 1.30 | | | | | | |
| log ε(Ce) | −0.51 | −0.16 | −0.02 | | | | | 0.75 | |
| [Ce/Fe] | 0.92 | 1.23 | 1.31 | | | | | 1.25 | |
| log ε(Pr) | −1.09 | −0.67 | −0.79 | | | | | | |
| [Pr/Fe] | 1.20 | 1.58 | 1.40 | | | | | | |
| log ε(Nd) | −0.21 | −0.01 | 0.25 | | 0.35 | | | 1.18 | |
| [Nd/Fe] | 1.38 | 1.54 | 1.74 | | 1.12 | | | 1.84 | |
| log ε(Sm) | −0.65 | −0.28 | −0.05 | | | | | | |
| [Sm/Fe] | 1.40 | 1.73 | 1.90 | | | | | | |
| log ε(Eu) | −0.81 | −0.71 | −0.52 | <−1.32 | −0.72 | −0.51 | <−0.62 | 0.21 | −0.14 |
| [Eu/Fe] | 1.68 | 1.74 | 1.87 | <1.50 | 0.95 | 1.70 | <2.40 | 1.77 | 2.11 |
| log ε(Gd) | −0.47 | −0.14 | | | | | | | |
| [Gd/Fe] | 1.47 | 1.76 | | | | | | | |
| log ε(Dy) | −0.29 | −0.15 | 0.20 | | 0.15 | 0.16 | | 1.22 | |
| [Dy/Fe] | 1.62 | 1.72 | 2.01 | | 1.24 | 1.79 | | 2.20 | |

**Extended Data Table 2:** Abundances of neutron-capture elements
Stars are numbered from brightest to faintest, from Star 1 to Star 9 in the order given by Extended Data Table 1.

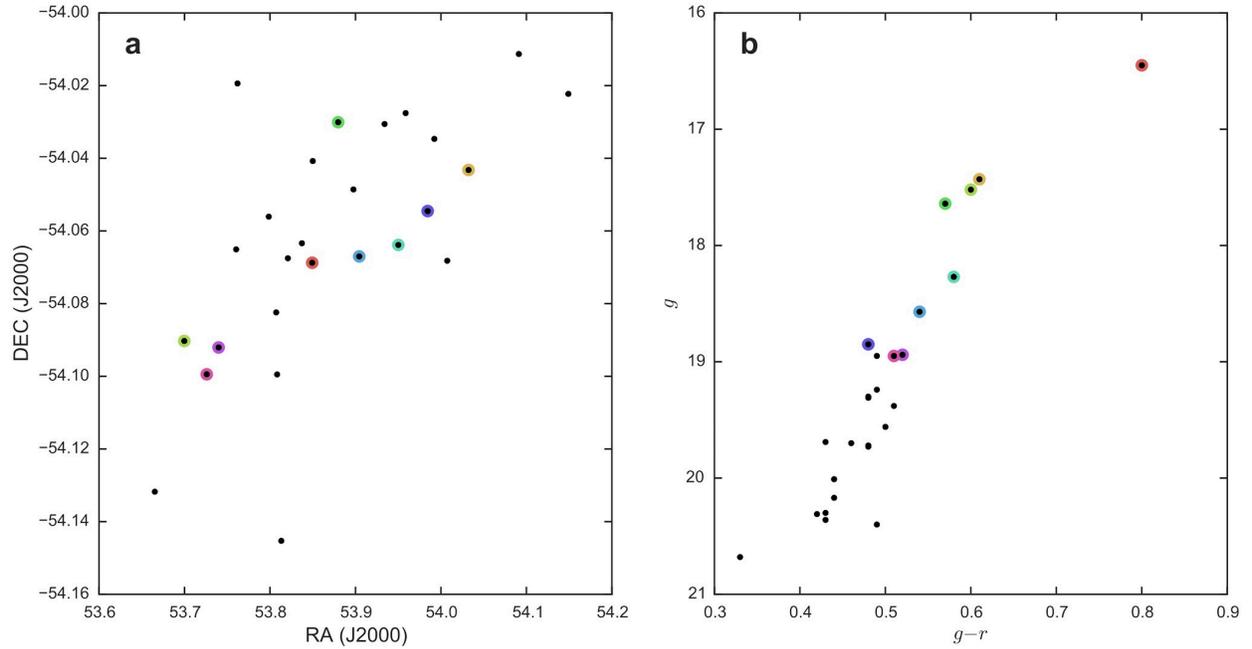

**Extended Data Figure 1:** Properties of Reticulum II member stars
**a**, Coordinates of member stars in Right Ascension and Declination[14]. Stars selected for observation with high-resolution spectroscopy are highlighted with large colored circles, while other members are shown in gray. **b**, Color-magnitude diagram based on Dark Energy Survey photometry[14].